\documentclass[submission]{eptcs}
\providecommand{\event}{SNR 2020} 
\usepackage{breakurl}             
\usepackage{underscore}           

\usepackage{amsthm}
\usepackage{amsmath}

\usepackage{microtype}
\usepackage{graphicx}
\usepackage{subfigure}
\usepackage{booktabs} 

\usepackage{amsfonts}
\usepackage{wrapfig}

\title{Analysis of E-commerce Ranking Signals \\ via Signal Temporal Logic}

\author{
	Tommaso Dreossi
	\institute{Amazon Search\\ Palo Alto, California, USA}
	\email{dreossit@amazon.com}
\and
	Giorgio Ballardin
	\institute{Amazon Search\\ Palo Alto, California, USA}
	\email{giobal@amazon.com}
\and
	Parth Gupta
	\institute{Amazon Search\\ Palo Alto, California, USA}
	\email{guptpart@amazon.com}
\and
	Jan Bakus
	\institute{Amazon Search\\ Palo Alto, California, USA}
	\email{jbakus@amazon.com}
\and
	Yu-Hsiang Lin
	\institute{Amazon Search\\ Palo Alto, California, USA}
	\email{yuhsianl@amazon.com}
\and
	Vamsi Salaka
	\institute{Amazon Search\\ Palo Alto, California, USA}
	\email{vsalaka@amazon.com}				
}

\def\titlerunning{Analysis of E-commerce Ranking Signals via Signal Temporal Logic}
\def\authorrunning{T. Dreossi, G. Ballardin, P. Gupta, J. Bakus , Y. Lin \& V. Salaka}
\begin{document}
\maketitle

\begin{abstract}
The timed position of documents retrieved by learning to
rank models can be seen as signals. Signals carry useful information
such as drop or rise of documents over time or user
behaviors. In this work, we propose to use the logic formalism
called Signal Temporal Logic (STL) to characterize document
behaviors in ranking accordingly to the specified formulas.
Our analysis shows that interesting document behaviors can
be easily formalized and detected thanks to STL formulas.
We validate our idea on a dataset of 100K product signals.
Through the presented framework, we uncover interesting
patterns, such as cold start, warm start, spikes, and
inspect how they affect our learning to ranks models.
\end{abstract}

\newcommand{\rob}[3]{\rho(#1, #2, #3)}
\newcommand{\todo}[1]{{\color{blue} TODO: #1}}
\newcommand{\apprx}{$\sim$}
\newtheorem{definition}{Definition}


\section{Introduction}

Learning to rank (LTR)~\cite{1500000016} is family of machine learning
techniques to solve ranking problems. Given a training set of
queries and documents sorted by some relevance degree, the
goal of an LTR model is to learn a model that, given a query,
sorts documents while maximizing the relevance score. The
relevance score can be either manually prepared from human
labelling or automatically derived from users interactions
logs. Automatic labelling is best suited for large amount
of data since it both relieves humans from the labelling
task and objectively measures the user preferences. In the
e-commerce context, clickthrough rate, that is the rate of the
clicks received by a retrieved product for a given query, is a
common example of automatic relevance score. In general, we
call behavioral signals the signals generated by user-ranker
interactions that can be used as relevance scores.
A drawback of using behavioral signals for training ranking
models is that products with low user interaction, such as
new or rare products, lack of behavioral signals and hence
are ranked as irrelevant. It takes time to gather enough
information so that the ranker can place the products in their
right position. This also leads to the causality dilemma: No
behavioral signals causes poor ranking which in turn results
in new products having a reduced likelihood of accruing
behavioral data. This particular phenomenon is referred to
as cold start which leads to a poor customer experience.

Cold start is just one particular problem rising from learning
from behavioral signals. Other examples are warm start
(product ranks too high too early), instability (sudden spikes
or ditches in ranking position), or uncertainty (the ranker
does not know how to rank due to lack of user interaction).
These are examples of well know unwanted phenomena that
an LTR model should avoid. Being able to isolate and measure
these phenomena plays an important role in designing
LTR rankers and preventing unwanted signal patterns.

Unluckily, there are few efficient tools for isolating known
signal patterns. Some examples are: probabilistic anomaly
detection methods~\cite{goldstein2016comparative}, where a signal is considered to
be an outlier accordingly to its probability of being observed;
k-means-based approaches~\cite{lima2010anomaly}, where the distance of
a signal from cluster centroids is a measure of diversity;
ad-hoc classification or regression models~\cite{steinwart2005classification}, where a
model is trained to detect a specific behavior. Note how these
techniques either do not provide the flexibility for identifying
a particular signal pattern or require to build rigid ad-hoc
solutions.

In this work we present use of Signal Temporal Logic
(STL)~\cite{maler2004}, as a tool for isolating and analyzing product
and behavioral signals in the LTR context. In particular,
we show how STL can be used to formally characterize well
known signal behaviors, such as the undesirable cold start,
warm start, product instability, etc., isolate them from a
collection of signals, and afterwards analyze them so that
we can take countermeasures in our LTR model. Intuitively,
STL is temporal logic~\cite{pnu77} (i.e., mathematical formalism
for representing and monitoring properties involving time)
particularly suitable for characterizing real-valued signals
defined over real-time intervals. Examples of signals patterns
in natural language that can be easily described by STL
formulas are \lq\lq some products always rank at position 1\rq\rq\ or
\lq\lq every product will eventually rank at position 1\rq\rq.

The main contributions of this work are:
\begin{itemize}
	\item Propose STL as a flexible tool for formally describing known signal patterns as formal logic formulas;
	\item Define a library of STL formulas encoding common unwanted signal behaviors in the LTR context (such as cold start, warm start, 	sudden ranking ditches or
spikes, instability, etc.);
	\item Cluster a large set of product signals collected from
a popular e-commerce website using the defined STL
properies and analyze how performance metrics, such
as clicks, impressions, and purchases are distributed
across different clusters and product categories;
	\item Compare the expressivity and succinctness of STL
with common signal clustering/filtering tools (such as
k-means, Pandas queries, propositional logics, etc.)
\end{itemize}

Researchers have widely used temporal logics for formal
verification purposes, where hardware and software systems
are tested against properties that characterize the system’s
correctness~\cite{fainekos2009robustness,jin2014powertrain}. Over the years, researchers
defined several types of temporal logics that usually vary
in expressive power. Some examples are Linear Temporal
Logic~\cite{pnu77}, Computation Tree Logic~\cite{clarke82}, or Metric
Temporal Logic~\cite{Koymans1990}. STL has been successfully applied
to the cyber-physical system domain~\cite{fainekos2009robustness,jin2014powertrain}, specifically
for monitoring and testing devices that involve physical
and computational components, such as drones~\cite{desai2017combining,pant2019fly},
self-driving vehicles~\cite{dreossi2019compositional,tuncali2018simulation}, and even
medical devices~\cite{bartocci2016computational}. The success of STL in these domains
is mainly due to its expressiveness and the efficiency of tools,
such as S-TaLiRo~\cite{annpureddy2011s} and Breach~\cite{donze2010breach}, for reasoning
with STL formulas. To the best of our knowledge, this is
the first time that STL is used in the context of Information
Retrieval and LTR rankers.

The paper is organized as follows. In Sec.~\ref{sec:stl} we explain
the theoretical promise of STL. In Sec.~\ref{sec:properties} we define a library
of STL properties useful for analyzing ranking signals. In
Sec.~\ref{sec:onstl} we also compare the expressiveness and succinctness
of STL to other common formalisms. In Sec.~\ref{sec:evaluation}, we present
experiments on evaluating and analyzing the defined STL
ranking properties on product signals collected from a popular
e-commerce website. Finally, we draw concluding remarks in
Sec.~\ref{sec:remarks}.


\section{Signal Temporal Logic}
\label{sec:stl}

In this section we define the Signal Temporal Logic~\cite{maler2004},
a formalism particularly suitable for properties of real-valued signals.

A signal is a function $s : D \to S$ with $D \subseteq \mathbb{R}_{\geq 0}$ an interval
and $S \subseteq \mathbb{R}$.
A trace $w = ( s_1, \dots, s_n )$ is a tuple of real-values signals defined over $D$.

Let $\Sigma = \{ \sigma_1, \dots, \sigma_k \}$ be a set of predicates $\sigma_i : \mathbb{R}^n \to \mathbb{B}$,
with $\sigma_i := p_i(x_1, \dots, x_n) \lhd 0$, $\lhd \in \{ <, \leq\}$, and $p_i : \mathbb{R}^n \to \mathbb{R}$.

\begin{definition}[STL syntax]
	An STL formula is defined by the grammar:
	\begin{equation}
		\varphi := \sigma | \neg \varphi | \varphi \wedge \varphi | \varphi U_{I} \varphi
	\end{equation}
	where $\sigma \in \Sigma$ and $I \subset \mathbb{R}_{\geq 0}$ is a closed non-singular interval.
\end{definition}

A shifted interval $I$ is defined as $t + I = \{ t + t' | t' \in I \}$.

\begin{definition}[STL qualitative semantics]
	Let $w$ be a trace, $t \in \mathbb{R}_{\geq 0}$, and $\varphi$ be an STL formula.
	The qualitative semantics of $\varphi$ is defined as follows:
	\begin{equation}
	\begin{aligned}
		& w, t \models \top \\
		& w, t \models p(x_1, \dots, x_n) \lhd 0 \text{ iff } p(w(t)) \lhd 0 \text{ with } \lhd \in \{<, \leq \} \\
		& w, t \models \neg \varphi \text{ iff } w, t \not\models \varphi \\
		& w, t \models \varphi_1 \wedge \varphi_2 \text{ iff } w, t \models \varphi_1 \text{ and } w, t \models \varphi_2 \\
		& w, t \models \varphi_1 U_I \varphi_2 \text{ iff } \exists t' \in t + I  \text{ s.t. } w, t' \models \varphi_2 \\ 
		& \qquad\qquad\qquad\qquad\qquad\qquad\qquad \text{ and } \forall t'' \in [t, t'] w, t'' \models \varphi_1 \\
	\end{aligned}	
	\end{equation}
\end{definition}

The peculiarity of STL is the interval-decorated until operator. Intuitively,
$\varphi_1 U_I \varphi_2$ holds if $\varphi_1$ is true until $\varphi_2$ 
becomes true at a time instant in $I$.

We can define other common operators as syntactic abbreviations:
$\bot := \neg \top ,
p(x) > 0 := \neg (p(x) \leq 0) ,
p(x) \geq 0 := \neg (p(x) < 0) ,
\varphi_1 \vee \varphi_2 := \neg ( \neg \varphi_1 \wedge \neg \varphi_2) ,
\varphi_1 \implies \varphi_2 := \neg \varphi_1 \vee \varphi_2 ,
F_I \varphi := \top U_I \varphi ,
G_I \varphi := \neg F_I \neg \varphi$.

The \lq\lq eventually\rq\rq\ operator $F_{I} \varphi$ (also called \lq\lq future\rq\rq) forces $\varphi$
to be true at least once in $I$. 
The \lq\lq always\rq\rq\ operator $G_{I} \varphi$ (also called \lq\lq globally\rq\rq) requires $\varphi$
to be always true in $I$.  
We will omit the interval decoration $I$ from temporal operators when the property
predicates over the entire life of trace (e.g., we write $G \varphi$ as a shorthand for $G_{[0, +\infty]} \varphi$).

We say that a trace $w$ satisfies $\varphi$ if $w, 0 \models \varphi$.
The satisfaction of a formula can be determined by recursively computing the satisfaction of its subformulas~\cite{maler2004}. The evaluation of an STL formula is linear
in the signal length~\cite{donze2013efficient}.

STL has also several alternative semantics. The most popular is the qualitative semantics that, instead of a boolean satisfaction value,
returns a real value encoding how robustly a trace satisfies a formula (i.e., the distance from satisfaction or violation). 
In this work, we consider only the qualitative semantics since we are interested in the binary classification
of our signals. For more details on qualitative semantics see, e.g.,~\cite{donze2013efficient}.

%
\section{Ranking Signal Properties}\label{sec:properties}

We now define three groups of specifications:
1) local properties, that predicate over parts of signals,
2) global properties, that involve whole signals, and
3) correctness properties, that monitor faulty behaviors such as missing data.

In the following experiments, we assume that the average
daily ranking position of a products is given by a function
$x : \mathbb{N} \to \mathbb{R}_{\geq 1}$, i.e., given a day $t_i \in \mathbb{N}$,
$x(t_i)$ is the daily average position of a document.
Let $x'(t_i)$ be the discrete-time derivative of $x$, i.e., $x'(t_i) = (x(t_{i+1}) - x(t_i))/(t_{i+1} - t_i)$.


\subsection{Local Properties \label{sec:localspecs}}

We begin with properties that predicate over the initial days
of products. We are interested in determining if products are ranked
at a stable position (flat start) or if they gain (cold start)
or loose (warm start) positions in the days after launch:

\begin{equation}
	flat\_start :=  G_{[0,w]}(|x'| < \epsilon ) 
	\label{eq:mtlflat}
\end{equation}
\begin{equation}
	cold\_start :=  G_{[0,w]}( x' \leq 0 ) \wedge F_{[0,w]}(x' < 0)
	\label{eq:mtlcold}
\end{equation}
\begin{equation}
	warm\_start :=  G_{[0,w]}( x' \geq 0 ) \wedge F_{[0,w]}(x' > 0)
	\label{eq:mtlwarm}
\end{equation}
where $w \in \mathbb{N}$ and $\epsilon \in \mathbb{R}_{\geq 0}$ are tunable
parameters that define the length of the initial time window
and noise tolerance.

The $flat\_start$ specification (Eq.~\ref{eq:mtlflat}) forces the first derivative $x'$
to be always $\epsilon$-close to zero (we include
the $\epsilon$ tolerance to account for noise). The ranking signals that
satisfy this specification are those that found their steady ranking position on launch day, 
i.e., those that experienced a flat start.
Similarly, we define $cold\_start$ (Eq.~\ref{eq:mtlcold}) and $warm\_start$ (Eq.~\ref{eq:mtlwarm})
specifications by requiring the ranking signals to always decrease/increase. The right-most
conjunct ($F_{[0,w]}$) forces the signals to strictly grow/decrease at least once.


\begin{figure}
	\centering  
	\subfigure[$flat\_start$ (Eq.~\ref{eq:mtlflat})]{\label{fig:flatstart}\includegraphics[width=0.32\textwidth]{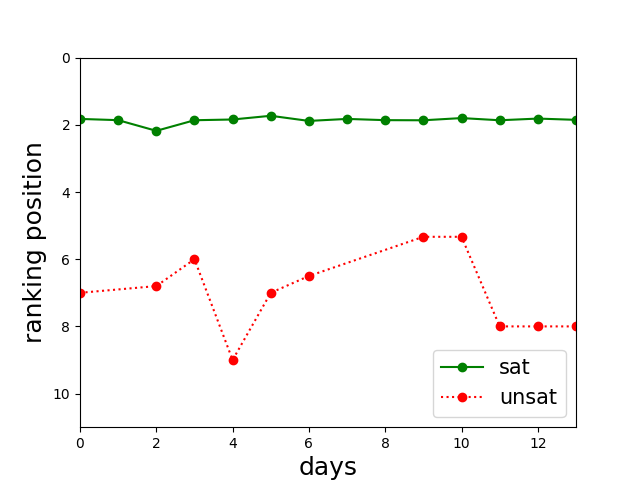}}
	\subfigure[$cold\_start$ (Eq.~\ref{eq:mtlcold})]{\label{fig:coldstart}\includegraphics[width=0.32\textwidth]{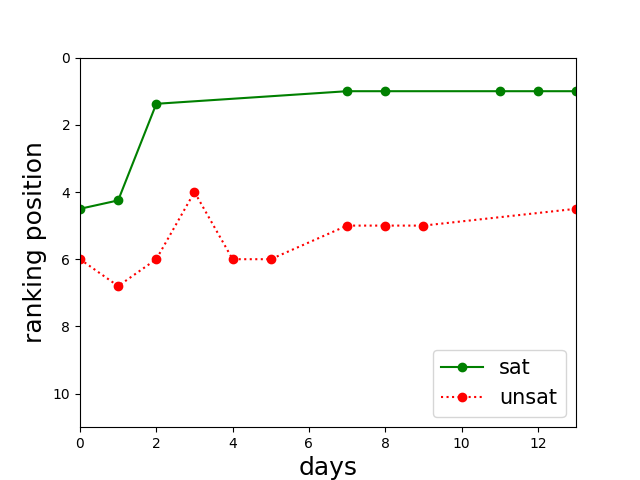}}
	\subfigure[$steady\_state$ (Eq.~\ref{eq:mtlsteady})]{\label{fig:steadystate}\includegraphics[width=0.32\textwidth]{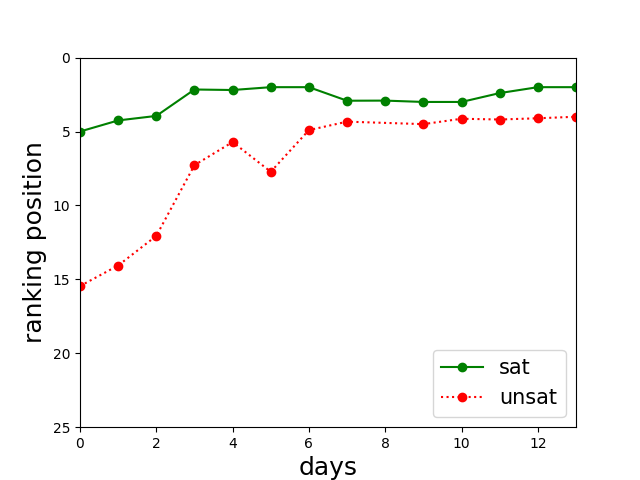}}
	\caption{Ranking signals that satisfy (green) and do not satisfy (red) the STL specifications $flat\_start$ (Eq.~\ref{eq:mtlflat}),
	$cold\_start$ (Eq.~\ref{eq:mtlcold}), and $steady\_state$ (Eq.~\ref{eq:mtlsteady}).\label{fig:starts}}
\end{figure}

Fig.~\ref{fig:starts} shows some examples of trajectories that do and do not satisfy (green and red, respectively)
the flat and cold start STL specifications with parameters $w = 3$ and $\epsilon = 1$ for $flat\_start$
and $w = 3$ and $\epsilon = 0$ for $cold\_start$.

\subsection{Global Properties}

We now increase the scope of our specifications by predicating over entire ranking signals.

In learning to rank, we often assume that a document reaches a steady state ranking position
after an initial period during which behavioral data is collected.
The following STL formula characterizes
the existence of a steady state:
\begin{equation}
	steady\_state:= F_{[0,w]} G( |x'| < \epsilon)
	\label{eq:mtlsteady}
\end{equation}
where $w \in \mathbb{R}_{\geq 0}$ defines the maximum stabilization time and
$\epsilon \in \mathbb{R}_{\geq 0}$ the tolerance to noise.
The $steady\_state$ formula holds if there is a day in $[0,w]$ after which $|x'|$ is always smaller than $\epsilon$,
i.e., the trajectory maintains a steady state with $\epsilon$ tolerance.
Fig.~\ref{fig:steadystate} shows two trajectories evaluated on $steady\_state$ with parameters $w = 3$ and $\epsilon = 1$.
The green signal satisfies the specification because it stabilizes after day $3$ at position $2$.
On the other hand, the red trajectory does not satisfy the specification because on day $4$ it experiences a drop 
in positions and thus it is not stable after day $3$.

\begin{figure}
	\centering  
	\subfigure[$reach$ (Eq.~\ref{eq:mtlreach})]{\label{fig:reach}\includegraphics[width=0.32\textwidth]{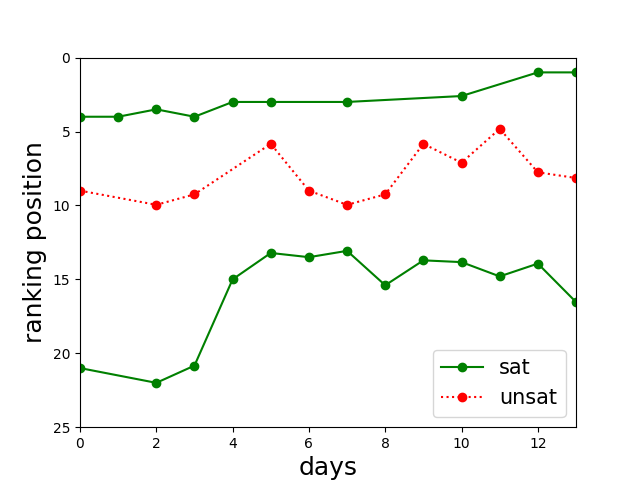}}
	\subfigure[$ditch$ (Eq.~\ref{eq:mtlditch})]{\label{fig:ditch}\includegraphics[width=0.32\textwidth]{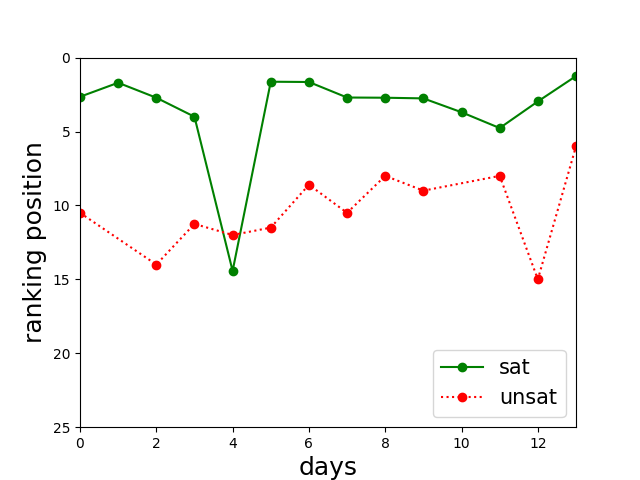}}
	\subfigure[$no\_long\_miss$ (Eq.~\ref{eq:mtlnolongmiss})]{\label{fig:nolongmiss}\includegraphics[width=0.32\textwidth]{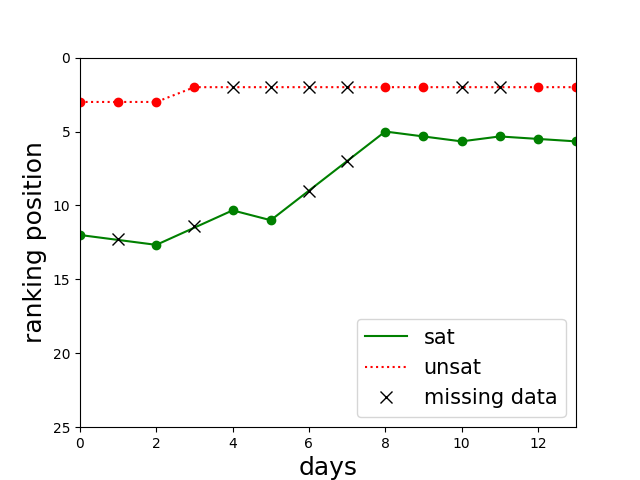}}
	\caption{Ranking signals that satisfy (green) and do not satisfy (red)
	the STL specifications $reach$ (Eq.~\ref{eq:mtlreach}), $ditch$ (Eq.~\ref{eq:mtlditch}), and $no\_long\_miss$ (Eq.~\ref{eq:mtlnolongmiss}).}
\end{figure}

Next, we define a liveness property that checks if a product that reaches a certain ranking interval eventually
hits a critical position:
\begin{equation}
	reach :=  G( (x < s) \implies F(x = r)  ) 
	\label{eq:mtlreach}
\end{equation}
with $s, r \in \mathbb{R}$. For instance, with parameters 
$s = 10$ and $r = 1$ we check whether products that reached the top $10$ positions eventually rank in first position too.
Fig.~\ref{fig:reach} depicts two signals that satisfy the $reach$ specification and one that does not with parameters $s = 10$ and $r = 1$.
The upper green signal satisfies the specification because it enters the $[1,10]$ ranking interval and eventually reaches position $1$ on day $12$.
The lower green signals also satisfies the requirement since it never enters the $[1,10]$ range. However, the red signal does not
satisfy the specification because its values become smaller than $10$ but never reach the first position.


Finally, we analyze the stability of products. 
We define two specifications that 
capture signals with large bouncing drops or rises in a short time frame:
\begin{equation}
	ditch :=  F( (x' > d) \wedge F_{[0,w]} (x' < d) ) 
	\label{eq:mtlditch}
\end{equation}
\begin{equation}
	spike :=  F( (x' < d) \wedge F_{[0,w]} (x' > d) ) 
	\label{eq:mtlspike}
\end{equation}

The parameters $d, w \in \mathbb{R}_{> 0}$ determine the signal drop/rise width and amplitude, respectively.
These two formulas check if at any point in time there is a drop/rise of at least $d$ positions
followed by a rise/drop of at least $d$ position within $w$ days.
Fig.~\ref{fig:ditch} shows some signals evaluated on $ditch$ with parameters $d = 10$ and $w = 2$.
The green signal satisfies the specification since it experiences a ditch of at least $10$ positions on day $3$.
The red signal does not satisfy the property since its ditch is not deep enough.

\subsection{Correctness Properties}\label{sec:correctnessspecs}

Finally, we hypothesize a scenario where some data might miss, i.e.,
the ranking position might be unknown. Let $-1$ denote
the ranking position of a product on a day for which data is missing.
The atomic predicate that holds if the ranking position is unknown is $miss := x = -1$.

In learning to rank systems, the first days after a product launch are crucial
for the collection of behavioral data and the subsequent correct ranking.
Hence, we do not want too much missing data in the
days after a product launch:
\begin{equation}\label{eq:mtlmiss}
	no\_init\_miss := \neg(G_{[0,w]} miss)
\end{equation}
where $w\in \mathbb{N}$ defines the initial time window.
$no\_init\_miss$ ensures that there are no $w$ consecutive initial days of missing data.
We can also extend this requirement to any part of a ranking signal and not just to its prefix.
We define a specification that ensures that are no windows with too many consecutive
days with missing data:
\begin{equation}\label{eq:mtlnolongmiss}
	no\_long\_miss := G (miss \implies F_{[0,w]} \neg miss)
\end{equation}
where $w\in \mathbb{N}$. $no\_long\_miss$ ensures that if there is a missing data day, then eventually 
within $w$ days there will be non missing data day.
In Fig.~\ref{fig:nolongmiss}, the red signal does not satisfy $no\_long\_miss$ for $w = 3$, since 
after day $4$ there are $4$ consecutive days of missing data. On the other hand, the green signal
satisfies the specification since it never has $3$ consecutive days of missing data.
 

\subsection{On STL's Succinctness and Efficiency \label{sec:onstl}}

\begin{wrapfigure}{L}{0.5\textwidth}
\centering
\includegraphics[width=0.45\textwidth]{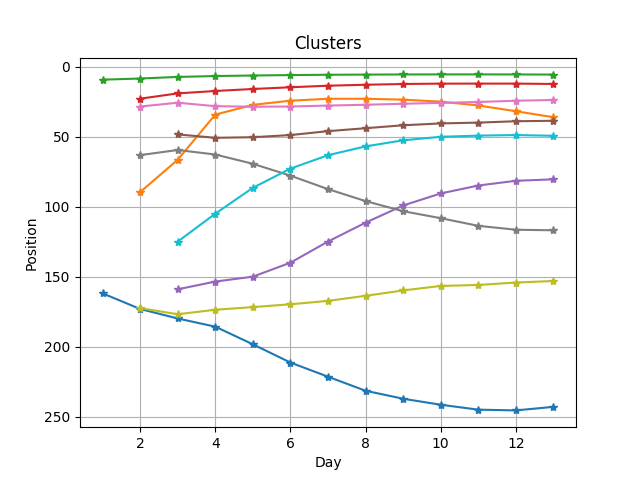}
\caption{k-means (with $k$= 10) centroid signals for
	our product ranking signals dataset. From this analysis
	it seems most products have smooth trajectories.
	However, our STL-based analysis reveals that more
	that 50\% of the signals experiences a spike or ditch
	of at least 10 ranking positions.\label{fig:kmeans}}
\end{wrapfigure}

Before proceeding with the evaluation of our STL specifications,
it is worth noticing how STL is both succinct and efficient
when compared with standard logics.

For instance, we ran the clustering algorithm k-means (with $k$=10) on our dataset
of product signals with the goal of mining the most representative
signals. Fig.~\ref{fig:kmeans} shows the centroid signals of the
clusters obtained by running k-means (with $k$=10) on product
ranking signals dataset. From this analysis it seems most
products have smooth trajectories. However, as we will later
discover in Sec.~\ref{sec:evaluation} , our STL-based analysis reveals that 50\%
of all the analyzed products in this test experience a shift in
ranking position over two consecutive days. This shows how
clustering techniques might fail in isolating important signal
patterns.


We also compared STL with classic logics and query formalisms.
For instance, the translation of the $ditch$ property (Eq.~\ref{eq:mtlditch})
in propositional logics is $\bigvee_{i=0}^T (xgd_i \wedge \bigvee_{j=0}^w xld_j)$
where $xgd_i := x'(i) > d$ and $xld_i := x'(i) < d$.
For this formulation, we must know in advance the length $T$ of the 
signal. In addition, it involves $T(1 + w)$ operators
against the $3$ operators of Eq.~\ref{eq:mtlditch}.

The encoding of Eq.~\ref{eq:mtlditch} in a first-order logic, e.g.,
semi-algebraic logics, would result in a formula with less operators
$\exists t ( x'(t) > d \wedge \exists t' (t \leq t' \leq t +w \wedge  x'(t') < d))$
which is still less compact than Eq.~\ref{eq:mtlditch} and less efficient to evaluate.
The evaluation of semi-algebraic formulas is generally doubly exponential 
in the number of quantifier alternations
while the evaluation of an STL formula is linear
in the signal length~\cite{donze2013efficient}.

For completeness, we also include the translation of Eq.~\ref{eq:mtlditch} into 
a \texttt{pandas}~\cite{mckinney-proc-scipy-2010} query, a Python library for data manipulation and analysis. 
Let $df$ be the dataframe with our signals where $pos_i$ is the column
with the ranking position on day $i$. The \texttt{pandas} query that isolates the signals that satisfy the
$ditch$ property is
$df[((df.pos_0 > d)\ \&\  (df.pos_1 < d\ |\ \dots \ | \ df.pos_w < d)) \ | \dots 
	|\ ((df.pos_{T-w} > d)\ \&\  (df.pos_{T-w+1}$ $ < d\ |\ \dots \ | \ df.pos_{T} < d))] \ $.
The structure of this query is similar to the propositional encoding.
Also in this case, we are dealing with a long query, we need to know in advance the length of the signal,
and any parametric change of our requirement affects the query's structure.
Finally, note that the encoding formula length explosion occurs for any specification that involves temporal
operators and not just this particular case.


\section{Experimental Evaluation}
\label{sec:evaluation}

We now evaluate the STL specifications defined in Sec.~\ref{sec:properties}
against a dataset of product signals. We conduct two analyses:
\begin{enumerate}
	\item \emph{Product categories}: Explore the correlation between query-product signals and product categories;
	\item \emph{Performance metrics}: Analyze how metrics such as impressions, clicks, and purchases distribute across different clusters of query-product signals.
\end{enumerate}

Our analyses highlight how STL can be used to easily isolate
unwanted signals behaviors. We will also discover that not
all product categories are equally affected by LTR anomalies
(e.g., cold start, instability, etc.) and that clicks, impressions,
and purchases are not evenly distributed across signal
patterns. Our dataset contains 100K examples from ten
different product categories. Each data point contains the product’s
position for 14 days, number of searches, clicks, and purchases.

For the evaluation of the specifications, we relied on the
\texttt{py-metric-temporal-logic library}~\cite{pyMTL}. Specifically, we
implemented a pandas~\cite{mckinney-proc-scipy-2010}  user-defined function that,
for each entry of our dataset, invokes \texttt{py-metric-temporal-logic} and evaluates our STL specifications.

\begin{figure}
	\centering  
	\subfigure[]{\label{fig:specrates}\includegraphics[width=0.49\textwidth]{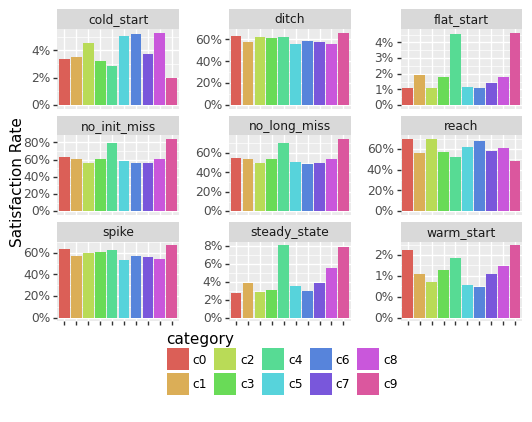}}
	\subfigure[]{\label{fig:metrics}\includegraphics[width=0.49\textwidth]{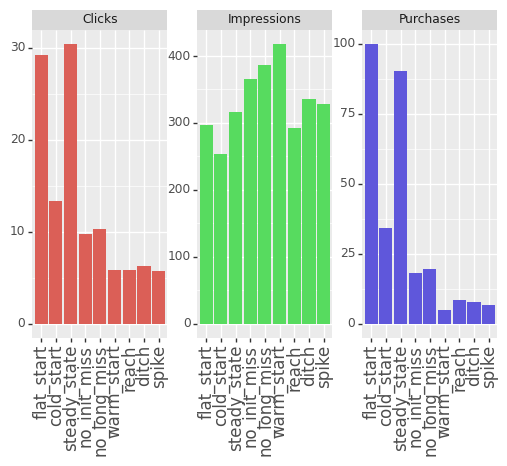}}
	\caption{Specification satisfaction rates across product
categories (left) and average total impressions, clicks, and purchases
garnered by query-products that satisfy STL
specifications (right).}
\end{figure}

\subsection{Product Categories}\label{sec:productcategories}

Do satisfaction rates of our STL specifications vary across
categories? To address this question we compute the percentage
of product signals that satisfies a given STL specification
for each product category and STL specification combination.
Results, reported in Fig.~\ref{fig:specrates}, highlight how satisfaction rates
are not equally distributed across different categories.

c9 and c4 are the categories less affected by cold start
with a 2\% satisfaction rate opposed to c8, c7, and c5 with
satisfaction rates 5\%. c4 and c9 have also the highest flat
start and steady state rates (5\% and 8\%, respectively) in
contrast to the other categories. This means that c4 and
c9 are the categories whose products most frequently find
their natural ranking position from launch day and keep it
constant over time.

Interestingly, c9, c4, together with c0, are the most affected
by warm start (1.50\%) suggesting that products from these
categories might rank too high after launch or might include
a high number of low engagements products (e.g., spam).

Finally, spike and ditch satisfaction rates are almost equally
distributed. Remarkably, 60\% of all the analyzed products
experience a drop or ditch of at least 10 ranking position
within two consecutive days.

\subsection{Performance Metrics}

We now analyze how impressions, clicks, and purchases are
distributed across different specifications. We compute the
average number of impressions, clicks, and purchases for
query-documents tuples that satisfy a given STL specification.
Fig.~\ref{fig:metrics} shows the obtained distributions.

The signals within this test that garner the highest and lowest
average number of impressions (400 and 250, respectively)
are those that satisfy the warm start and cold start
specifications, respectively. The most clicked products satisfy
flat start and steady start (30 clicks) while the least
clicked ones satisfy warm start, reach, ditch, and spike (5
clicks). The products that collect highest purchases satisfy
flat start and steady state (90) followed by cold start
(35), while the products with lowest purchases are those
affected by warm start (5).

This analysis suggests that products affected by warm
start, despite receiving the highest number of impressions,
tend not to be clicked and consequently gather low purchases.
We could speculate that warm start products are either not
relevant to customer’s searches or are low quality products
that do not attract customer’s attention. cold start products
show the symmetric phenomenon. They receive a low number
of impressions, they are clicked twice as much as warm start
and gather 5x purchases compared to warm start. This might
mean that cold start products are eventually discovered and
purchased by customer despite being initially poorly ranked.

Outliers are flat start and steady state specifications.
They receive 3x clicks (30 vs 10) and 4x purchases (100
vs 25) compared to the second mostly clicked and highest
purchases cold start, no init miss, and no long miss. Products
that have early flat and steady positions tend to rank
very high and consequently collect the highest amount of
clicks and purchases.

Note how both the product categories and performance
metrics STL analyses revealed interesting behaviors of our
learning to rank model. The obtained insights can be used, for instance,
to rebalance our training sets by focusing on particular product
segments or redesign relevance scores. Note also that
these are just demonstrative examples of how STL can be
used to reason over ranking signals. Nothing prevents STL
from being applied to more complex temporal properties or
more sophisticated analyses.

\section{Remarks}
\label{sec:remarks}

In this work, we proposed for the first time STL in the learning to rank
context to cluster and analyze product ranking signals. We
defined a library of properties that characterize unwanted
product behaviors and analyzed the distribution of the satisfaction
of properties over a dataset of 100K product signals.
Our analysis showed how STL can be used to reason over
ranking traces and reveal insights on the model under study.
In the future, we plan to explore STL for online monitoring
where the real-time detection of faulty behaviors can be used
to trigger alarms and actuate repairing procedures.


\bibliographystyle{eptcs}
\bibliography{biblio}
\end{document}